\documentclass[twocolumn,floatfix,prb,aps,showpacs]{revtex4}
\usepackage{graphicx,amsmath,amssymb}

\begin{document}

\newcommand{\CORR}[1]{#1}

\title{States of interacting composite fermions at Landau level fillig $\nu=2+3/8$}

\author{Csaba T\H oke$^{1,2}$, Chuntai Shi$^3$, and Jainendra K.~Jain$^{3}$}
\affiliation{$^{1}$Physics Department, Lancaster University, LA1 4YB, Lancaster, United Kingdom}
\affiliation{$^{2}$Institut f\"ur Theoretische Physik, Johann Wolfgang Goethe-Universit\"at, 60438 Frankfurt/Main, Germany}
\affiliation{$^{3}$Department of Physics, 104 Davey Lab, Pennsylvania State University, University Park PA, 16802}
\date{\today}

\begin{abstract}
There is increasing experimental evidence for fractional quantum Hall effect at filling factor $\nu=2+3/8$.  Modeling it as a system of composite fermions, we study the problem of interacting composite fermions by a number of methods.  In our variational study, we consider the Fermi sea, the Pfaffian paired state, and bubble and stripe phases of composite fermions, and find that the Fermi sea state is favored for a wide range of transverse thickness.  However, when we incorporate interactions between composite fermions through composite-fermion diagonalization on systems with up to 25 composite fermions, we find that a gap opens at the Fermi level, suggesting that inter-composite fermion interaction can induce fractional quantum Hall effect at $\nu=2+3/8$. The resulting state is seen to be distinct from the Pfaffian wave function. 
\end{abstract}

\maketitle

\section{Introduction}

The fractional quantum Hall effect \cite{Tsui} (FQHE) is understood as a consequence of the formation of bound states of electrons and quantized vortices, known as composite fermions\cite{Jain}.  In particular, the FQHE at fractions belonging to the sequences $\nu=s\pm\frac{n}{2pn\pm1}$ is a manifestation of the integral quantum Hall effect (IQHE) of composite fermions at the composite-fermion filling $\nu^*=n$.  In recent years, the FQHE at fractions not belonging to the these sequences 
has attracted interest, because it cannot be explained in terms of a 
model of {\em non}interacting composite fermions, which only exhibit IQHE.  
In many instances, these new fractions can be shown to arise from the weak residual  interactions between  composite fermions.
For example, the FQHE at $\nu=5/2$ is understood as a p-wave Pfaffian-paired state of composite fermion\cite{exp5p2,theory5p2},
and the FQHE at 4/11 as a {\em fractional} QHE of composite fermions\cite{Pan1,Chang,Quinn}.
Recent experiments in very high mobility samples have found signatures \cite{Pan2,Kang} for FQHE at $\nu=2+3/8$,
albeit with a tiny gap of a few mK.
Although the evidence is not yet conclusive, the possibility of this FQHE is particularly exciting both because it
is an even denominator fraction and because this fraction occurs in the second Landau level,
where FQHE is not as extensive as in the lowest Landau level.
That has motivated us to examine various theoretical possibilities at this filling factor proceeding with the assumption that
the second Landau level 3/8 state can be modeled as composite fermions at filling 3/2.
Within this model, we rule out several simple variational states, but find an instability of the
composite fermion Fermi sea into a gapped state; while the true nature of this FQHE state is not fully understood at present, our studies indicate that it is distinct from the usual Pfaffian state.  We demonstrate the robustness of this state against finite thickness.

The strongly interacting system of electrons confined to a Landau level is described in terms of exotic emergent particles called composite fermions, which are bound
states of an electron and an even number ($2p$) of vortices of the many-body wave function. The most dramatic consequence of the composite fermion (CF) formation is that the Berry phase due to the attached vortices effectively cancels
part of the external magnetic field, producing dynamics governed by a reduced field $B^\ast=B-2p\rho\phi_0$,
where $\rho$ is the CF density and $\phi_0=hc/e$ is the flux quantum.
Composite fermions form Landau-like levels, called $\Lambda$ levels, with their filling factor related to the electronic
filling factor through the expression $\nu=\frac{\nu^\ast}{2p\nu^\ast\pm 1}$.
The CF formation correctly accounts for most of the correlation effects, and a model of weakly interacting composite
fermions securely explains the prominent experimental observations, including the FQHE $\nu=\frac{n}{2pn\pm1}$ as the IQHE
of composite fermions\cite{Jain} and the compressible liquid at $\nu=1/(2p)$ as the Fermi sea of composite fermions\cite{HLR}.
More subtle structures can emerge due to the weak residual interactions between composite fermions\cite{theory5p2,newodd,paper3,Lee,CFint,Mandal,Wojs2}.

Treating the lowest filled Landau level as inert, the problem of our interest is that of interacting electrons in the second LL at $\nu^{(1)}=3/8$.  
The dimension of the Hilbert space here is too large to obtain meaningful results from exact diagonalization.
We proceed instead  by modeling $\nu^{(1)}=3/8$ as a state of composite fermions at filling factor $\nu^\ast=3/2$.
Assuming that composite fermions are fully spin polarized, the state contains a fully occupied lowest $\Lambda$ level,
and a half filled second $\Lambda$ level.
We assume that the lowest $\Lambda$ level is inert and work with only the composite fermions of the half filled second
$\Lambda$ level in the rest of the paper, denoting their number by $N$.

Several states of composite fermions have been considered at a half filled Landau level:
Fermi sea, paired Pfaffian state, stripes, and bubble crystals.  When considering the first two states at half-filled {\em second} $\Lambda$ level,
the $^2$CFs in the second $\Lambda$ level capture two additional vortices to transform into $^4$CFs,
thereby producing very complex mixed structures.
(The symbol $^{2p}$CF refers to composite fermions carrying $2p$ vortices.)
Which, if any, of these states is stabilized in nature depends on the residual interaction between composite fermions
in their second $\Lambda$ level, which itself is a remnant of the Coulomb interaction between electrons occupying the second Landau level.

It is interesting to note that when composite fermions in the second $\Lambda$-level capture an additional pair of vortices, as is the case for the CF-Fermi sea or the Pfaffian state, {\em three} species of fermions coexist in the system:
electrons in the lowest LL, $^2$CFs in the lowest $\Lambda$-level of the second LL, and and $^4$CFs in the second $\Lambda$-level; only the last will be explicitly considered in our calculations.  Much or our effort will be toward obtaining the effective interaction between them by integrating out the remaining fermions.

\CORR{Our paper is organized as follows.
In Sec.~\ref{intercf} an effective inter-CF interaction is derived for composite fermions in the second $\Lambda$ level of the second Landau level; 
both pseudopotential and real-space representations are obtained.
Using this interaction several variational states are compared energetically in Sec.~\ref{varist}.
Exact diagonalization results in Sec.~\ref{exact} confirm the relative advantage of the CF Fermi sea state.
Then, in Sec.~\ref{cfdiag}, the residual interaction between composite fermions is included perturbatively
to explore any further instability of the CF Fermi sea state. Sec.~\ref{concl} summarizes  the principal conclusions of our study.}

\section{Inter-composite-fermion interaction}
\label{intercf}

The determination of the inter-CF interaction proceeds along several steps.
First of all, following standard practice, we represent the second LL Coulomb interaction
(including finite thickness effects) as an effective interaction in the lowest LL (with zero thickness),
for which we use the form\cite{Shi}
\begin{eqnarray}
V^{\text{eff}}(r)&=&\frac{1}{r}+\frac{B_3}{\sqrt{r^6+A_3}}+\frac{B_5}{\sqrt{r^{10}+A_5}}+\frac{B_7}{\sqrt{r^{14}+A_7}}+\notag\\
&&\sum_{i=0}^3 C_i r^i e^{-r^2},
\label{second}
\end{eqnarray}
where the constants $B_i,C_i$ are evaluated by matching the first few pseudopotentials for the two problems.
This form is motivated by the following observations:
(i) The Fourier transform of the exact effective interaction for $w=0$ is\cite{Haldane,Ambru} $V^{\rm eff}(q)=\frac{2\pi}{q}\left(1-\frac{q^2}{2}\right)^2$. Its inverse Fourier
transform $V^{(1)}(r)=\frac{1}{r}+\frac{1}{r^3}+\frac{9}{4r^5}$, however, is ill-behaved in that it yields divergent pseudopotentials for relative angular momenta $m=0,1$.
(ii) Regularizing the interaction as in Eq.~(\ref{second}) removes the short distance  divergences without significantly altering the long-distance behavior. 
(iii) Adding short-range Gaussian terms and fitting the first few pseudopotentials takes care of the short range part of the interaction without affecting the long-distance behavior.   The constants $A_j$ are arbitrary; we choose $A_3=1$, $A_5=10$, $A_7=100$ to maximize the efficiency of our calculation.  The interaction in Eq.~(\ref{second}) reproduces all second LL pseudopotentials almost exactly.
\CORR{The transverse thickness $w$ is modeled through a square quantum well potential with the electronic wave function given by
$\psi(z)=\sqrt\frac{w}{2}\cos\left(\frac{z\pi}{w}\right)$ in the transverse dimension.
%%%%For further details see Ref.~\onlinecite{Shi}.
Table \ref{coeff} gives $B_i,C_i$ for various values of transverse thickness $w$,
as well as the greatest relative error in pseudopotentials due to the approximations made in the form of Eq.~(\ref{second}).}

\begin{table*}[htb]
\begin{center}
\begin{tabular}{c|c|c|c|c|c|c|c|l}
\hline\hline
$w/\ell_B$ & $B_3$ & $B_5$ & $B_7$ & $C_0$ & $C_1$ & $C_2$ & $C_3$ & rel. error at $m=9$\\
\hline
0   & 1       &  2.25    &  0       & -29.6652  & 25.9333 &  -5.7924  & 0.35502  & $-5\times 10^{-6}$ \\
0.6 & 0.98824 &  2.1447  & -0.64764 & -26.7113  & 22.8682 &  -5.05705 & 0.308495 & $-4\times 10^{-6}$ \\
1   & 0.96733 &  1.96027 & -1.73059 & -23.7501  & 20.1878 &  -4.46304 & 0.27412  & $-3\times 10^{-6}$ \\
1.2 & 0.95295 &  1.83554 & -2.42498 & -22.4375  & 19.2136 &  -4.2825  & 0.266517 & $-1\times 10^{-6}$ \\
1.8 & 0.89414 &  1.3427  & -4.85138 & -19.8480  & 18.4978 &  -4.41921 & 0.299473 & $ 7\times 10^{-6}$ \\
2   & 0.86931 &  1.14304 & -5.68157 & -19.53315 & 19.0897 &  -4.72306 & 0.332065 & $ 1\times 10^{-5}$ \\
2.4 & 0.81181 &  0.69985 & -7.17964 & -19.7498  & 21.4679 &  -5.70842 & 0.429353 & $ 2\times 10^{-5}$ \\
3   & 0.70595 & -0.04589 & -8.4768  & -21.80525 & 27.2074 &  -7.87814 & 0.634219 & $ 5\times 10^{-5}$ \\
4   & 0.47724 & -1.3468  & -5.64371 & -25.4366  & 34.7657 & -10.7723  & 0.907296 & $ 9\times 10^{-5}$ \\
5   & 0.18318 & -2.39612 &  6.65526 & -17.5712  & 20.2936 &  -6.19823 & 0.513039 & $ 6\times 10^{-5}$ \\
\hline\hline
\end{tabular}
\end{center}
\caption{\label{coeff}
The coefficients of the second Landau level effective interaction (Eq.~(\ref{second})) as a function of quantum well thickness.
}
\end{table*}

\begin{figure}[!htbp]
\begin{center}
\includegraphics[width=\columnwidth, keepaspectratio]{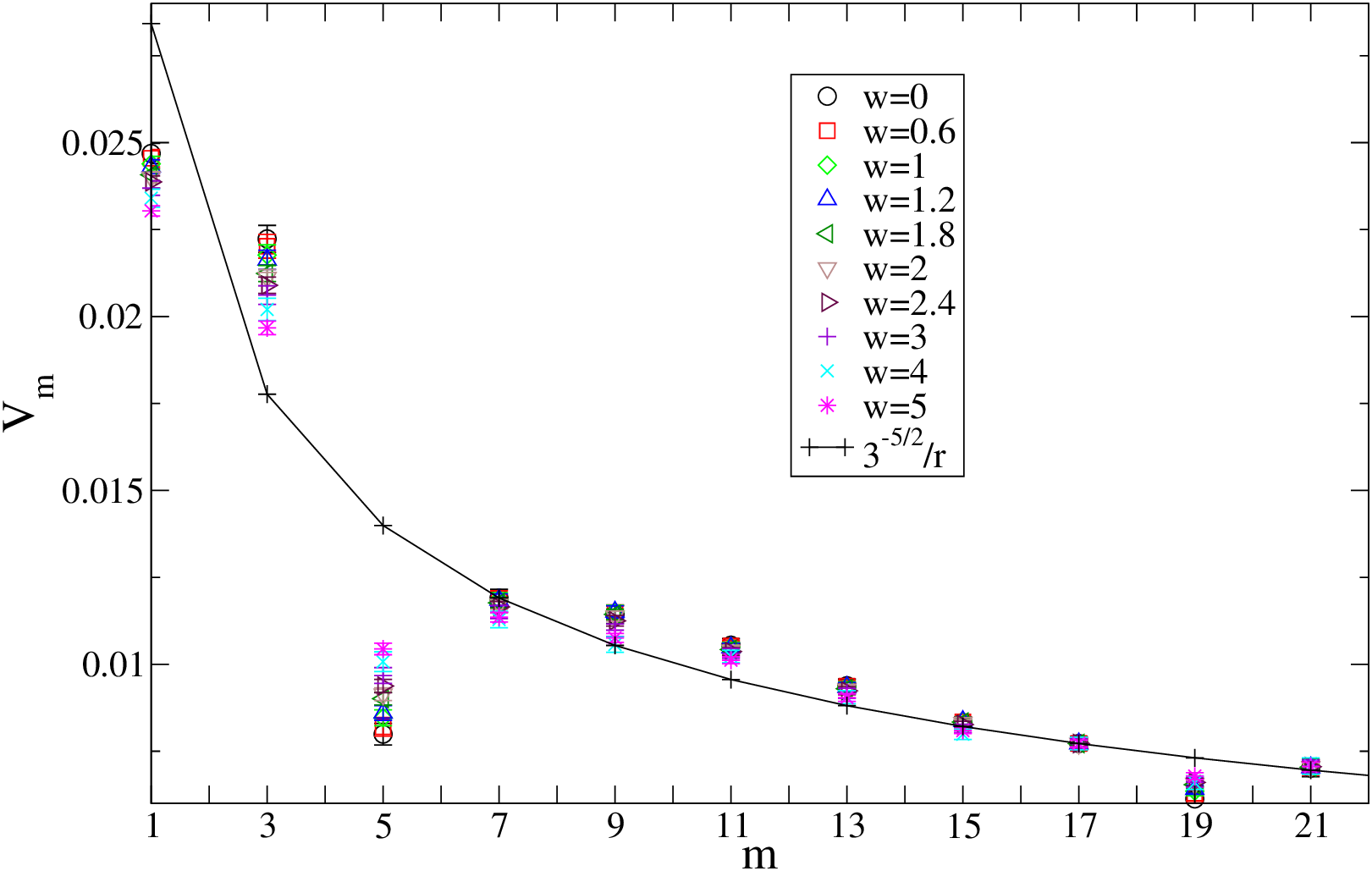}
\end{center}
\caption{\label{ppcomp} (Color online)
The pseudopotentials for two composite fermion quasiparticles in the $n=1$ LL, and
the asymptotic form $(2n+1)^{-5/2}V^{(0)}_m=3^{-5/2}\frac{\Gamma(m+1/2)}{2\Gamma(m+1)}$.
\CORR{The solid line on the asymptotic form is a guide to the eye.}
}
\end{figure}

The next step is to determine the interaction pseudopotentials\cite{Haldane} for composite fermions in the second $\Lambda$-level 
following Refs.~\onlinecite{Lee,Wojs}, by evaluating the energy of the state with two composite fermions in relative angular momentum $m$ state,
the wave function for which can be constructed explicitly according to the standard CF theory\cite{Jain}.
\CORR{
(Another formalism\cite{murthy} has also been used for a treatment of the inter-CF interactions at the  Hartree-Fock level.
However, that approach is designed for long distance physics and is not reliable for absolute energy comparisons of competing CF states.)
Our calculation is based on the Monte Carlo method in the spherical geometry\cite{Haldane},
in which electrons move on the surface of a sphere and a radial magnetic field is produced by a magnetic monopole of strength $Q$ at the center.
Here $2Q\phi_0$ is the magnetic flux through the surface of the sphere; $\phi_0=hc/e$; and $2Q$ is an integer according to Dirac's quantization condition.
The single particle states are monopole harmonics\cite{Wu} $Y_{Qlm}$, where
$l=Q+n$ is the angular momentum with $n=0,1,\ldots$ being the LL index,
$m=-l,-l+1,\ldots,l$ is the $z$-component of angular momentum.
Composite fermion states are defined as\cite{Jain,JainKamilla}
\begin{equation}
\label{cfstate}
\Psi^{\text{CF}}={\cal P_{\rm LLL}}\prod_{i<j}(u_iv_j-v_iu_j)^2\Phi,
\end{equation}
where $u\equiv\cos(\theta/2)\exp (-i\phi /2)$, $v\equiv\sin(\theta/2)\exp(i\phi /2)$, $\Phi$ is a Slater determinant of $Y_{Qlm}$'s,
and ${\cal P_{\rm LLL}}$ is the lowest LL projection.
(The planar geometry equivalents are obtained, apart from the Gaussian factor $e^{-\frac{1}{4}\sum_i|z_i|^2}$, by the substitution $(u_iv_j-v_iu_j)\iff(z_i-z_j)$, where $z_i=x_i-iy_i$ denotes the coordinates of the $i$th particle on the plane.)
For the pseudopotentials $V_m^{\text{CF}}$ in the second $\Lambda$ level, we consider states that contain two composite fermions
in the second $\Lambda$ level above a fully occupied lowest $\Lambda$ level.
In the spherical geometry the pseudopotentials are size, or $N$ dependent.  For any specific $N$ the pseudopotential $V^{(N)}_m$ is the energy of two composite fermions at relative angular momentum $m$.
The interaction (Eq.~(\ref{second})) is evaluated for such states assuming that $r$ in Eq. 1 is the chord distance.
%%%%For an $N$-particle system on the sphere, the pseudopotential $V^{(N)}_m$ is the energy of the state with total angular momentum $2Q-m$.
To allow for a comparison between systems with different numbers of particles, an additive constant is chosen 
to fit the largest $m$ pseudopotential to the expected asymptotic value $V_m=3^{-5/2}\frac{\Gamma(m+1/2)}{2\Gamma(m+1)}$
between point-like charge $e/3$ objects at long distances.
(The prefactor accounts for both the fractional charge and the conversion factor for energies expressed in terms of $\frac{e^2}{\epsilon\ell_B}$ and
$\frac{e^2}{\epsilon\ell^\ast_B}$, where $\ell^\ast_B=\sqrt3\ell_B$ is the magnetic length for composite fermions\cite{Lee} and $\ell_B=\sqrt{\hbar c/eB}$.)
Then $V_m^{\text{CF}}$ is obtained from an linear extrapolation to the $N\to\infty$ limit
of $V^{(N)}_m$ as a function of $1/N$. We used $18\le N\le50$ for this extrapolation.}
As seen in Fig.~\ref{ppcomp}, the inter-CF interaction \CORR{smoothly connects to} the interaction between point-like charge $e/3$ objects
%%%%%%$V_m=3^{-5/2}\frac{\Gamma(m+1/2)}{2\Gamma(m+1)}$,
at long distances, with significant deviation at short distances.
The assymptotic expression is used for $m\ge17$, where the numerically calculated pseudopotential is fitted to the assymptotics modulo an additive constant.
As expected, the transverse thickness weakens the short-range part of the interaction.

Armed with the CF pseudopotentials, we finally map the system into that of fermions at $\nu'=\nu^\ast-1=1/2$ in the {\em lowest} Landau level.  To the pseudopotentials of Fig.~(\ref{ppcomp}) an effective real-space interaction can be associated, for which we use the form\cite{Lee}
\begin{equation}
\label{effective}
V^{(w)}(r^\ast)=\left(\sum_{i=0}^5 c^{(w)}_{8i+4}{r^\ast}^{8i+4}e^{-{r^\ast}^2}+\frac{(2n+1)^{-5/2}}{r^\ast}\right)\frac{e^2}{\epsilon\ell_B},
\end{equation}
where $r^\ast$ is the distance measured in units of $\ell^\ast_B$.
With six parameters (Table \ref{icfcoeff}), all odd pseudopotentials $V_m$ from $m=1$ to $m=13$ can be fitted exactly,
and Eq.~(\ref{effective}) already has the correct long-range behavior.

\begin{figure}[!htbp]
\begin{center}
\includegraphics[width=\columnwidth, keepaspectratio]{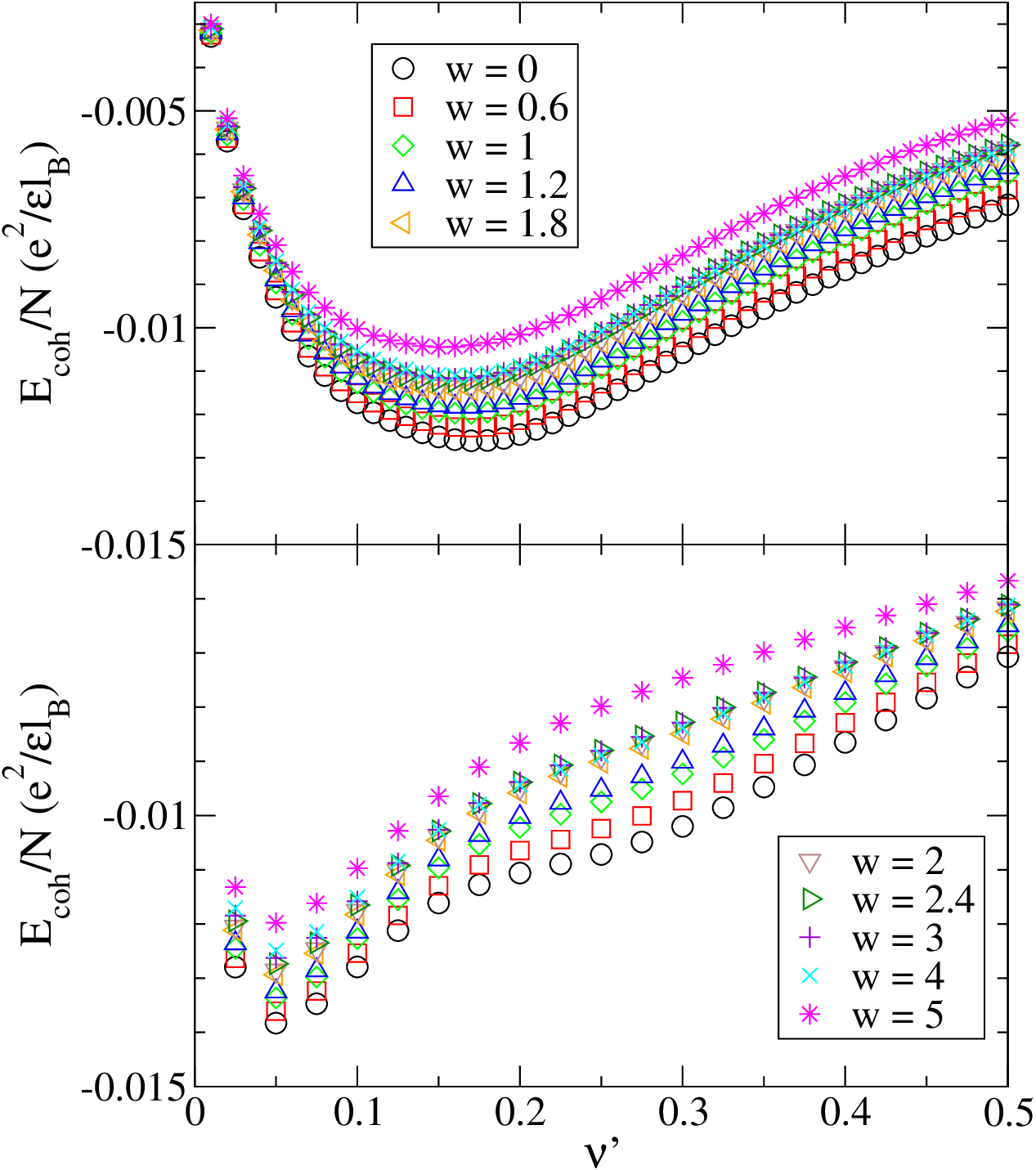}
\end{center}
\caption{\label{bs} (Color online)
Cohesive energy for the stripe (top) and bubble (bottom) phases of composite fermion quasiparticles in the second $\Lambda$-level of second Landau level.
}
\end{figure}

\begin{table*}[htb]
\begin{center}
\begin{tabular}{c|c|c|c|c|c|c}
\hline\hline
$w/\ell_B$ & $c_4$ & $c_{12}$ & $c_{20}$ & $c_{28}$ & $c_{36}$ & $c_{44}$\\
\hline
0   & 
    -0.275427217 &
    0.00317402221 &
    -1.52988973$\times 10^{-6}$ &
    9.54557755$\times 10^{-11}$ &
    -1.15955346$\times 10^{-15}$ &
    2.71593990$\times 10^{-21}$ \\
0.6 & 
    -0.264914310 &
    0.00302703755 &
    -1.45184249$\times 10^{-6}$ &
    9.00928816$\times 10^{-11}$ &
    -1.088766805$\times 10^{-15}$ &
    2.540345047$\times 10^{-21}$ \\
1   & 
    -0.252523057 &
    0.00285573396 &
    -1.36226008$\times 10^{-6}$ &
    8.40500455$\times 10^{-11}$ &
    -1.01043093$\times 10^{-15}$ &
    2.34859495$\times 10^{-21}$ \\
1.2 & 
    -0.246070821 &
    0.00276724752 &
    -1.31658053$\times 10^{-6}$ &
    8.10191825$\times 10^{-11}$ &
    -9.71797377$\times 10^{-16}$ &
    2.25524476$\times 10^{-21}$ \\
1.8 & 
    -0.227976782 &
    0.00252196654 &
    -1.19271759$\times 10^{-6}$ &
    7.30411893$\times 10^{-11}$ &
    -8.73307438$\times 10^{-16}$ &
    2.02334493$\times 10^{-21}$ \\
2   & 
    -0.222696641 &
    0.00245136596 &
    -1.15814270$\times 10^{-6}$ &
    7.09114678$\times 10^{-11}$ &
    -8.48361037$\times 10^{-16}$ &
    1.96725999$\times 10^{-21}$ \\
2.4 & 
    -0.213492009 &
    0.00232970992 &
    -1.10039082$\times 10^{-6}$ &
    6.75254404$\times 10^{-11}$ &
    -8.11190196$\times 10^{-16}$ &
    1.88888211$\times 10^{-21}$ \\
3   & 
    -0.202852166 &
    0.00219192928 &
    -1.03924014$\times 10^{-6}$ &
    6.43520873$\times 10^{-11}$ &
    -7.82651887$\times 10^{-16}$ &
    1.84268458$\times 10^{-21}$ \\
4   & 
    -0.189229729 &
    0.00201494874 &
    -9.63279105$\times 10^{-7}$ &
    6.06471130$\times 10^{-11}$ &
    -7.53214883$\times 10^{-16}$ &
    1.80500175$\times 10^{-21}$ \\
5   & 
    -0.168823778 &
    0.00172511349 &
    -8.08579450$\times 10^{-7}$ &
    4.97487537$\times 10^{-11}$ &
    -6.04378114$\times 10^{-16}$ &
    1.42355148$\times 10^{-21}$ \\
\hline\hline
\end{tabular}
\end{center}
\caption{\label{icfcoeff}
The coefficients of the effective inter-CF interaction in the second $\Lambda$ level of the second Lnadau level
(Eq.~(\ref{effective})) as a function of quantum well thickness.
}
\end{table*}

\section{Variational states}
\label{varist}

We first consider charge-density-wave states of composite fermions, both stripes and bubble crystals; analogous states have proved to be relevant for half-filled electronic LLs with high LL index\cite{Koulakov}.  (We note that liquid crystalline phases of electrons, possibly with nematic order, have also been considered in high Landau levels\cite{liquidcrystal}; we do not consider in this work analogous phases for composite fermions.)
In the Hartee-Fock scheme the cohesive energy of these states is\cite{Koulakov}
\begin{equation}
\label{coh}
E_{\text{coh}}=\frac{(2\pi)^3}{2NL_xL_y}\sum_{\mathbf q\neq0}\tilde U_{\text{HF}}(q)\Delta(-\mathbf q)\Delta(\mathbf q),
\end{equation}
which is defined as the interaction energy measured from the uniform Hartree-Fock state
\begin{equation}
\label{uniform}
E_0=-\frac{\tilde U(q=0)}{2}\nu'.
\end{equation}
\CORR{This expression is based on the assumption that the CF-background and the background-background interaction also have the
same form as the CF-CF interaction in Eq.~(\ref{effective}); because the first two are identical for all uniform states, their actual form is not relevant to the energy comparisons, and can be chosen according to convenience.}
The quantity $\Delta(\mathbf q)=\frac{1}{2\pi}\sum_ke^{-kq_x(\ell^\ast_B)^2}\langle a^\dag_{k_+} a_{k_-}\rangle$ in Eq.~(\ref{coh}) is the orbit-center density, and we define 
$\tilde U_{\text{HF}}(q)=\tilde U(q)-(\ell^\ast_B)^2U(q\ell^\ast_B)$,
$\tilde U(q)=V^{(w)}(q)e^{-\frac{1}{2}q^2(\ell^\ast_B)^2}$, and $k_{\pm}=k\mp q_y/2$.
For the stripe phase this reduces to
\begin{equation}
E_{\text{coh}}^{\text{stripe}}=\frac{1}{2\nu'(\ell^\ast_B)^2}\sum_{q\neq0}\tilde U_{\text{HF}}(q)\left(
\frac{2\sin\frac{q\Lambda_s\nu'}{2}}{\Lambda_s q}\right)^2,
\end{equation}
where $q=\frac{2j\pi}{\Lambda_s}$, and for the bubble crystal
\begin{equation}
E_{\text{coh}}^{\text{bubble}}=\frac{2\pi^2 l_0^2}{\nu'}\sum_{\mathbf q\neq 0}\tilde U_{\text{HF}}(q)
\left(\frac{R}{Al_0^2q}J_1(qR)\right)^2,
\end{equation}
with $R=\Lambda_b\sqrt{\sqrt3 \nu'/2\pi}$, $A=(\sqrt3/2)\Lambda_b^2$, and $\mathbf q=n\mathbf e_1+m\mathbf e_2$ with $\mathbf e_1=\frac{4\pi}{\sqrt3\Lambda_b}\mathbf{\hat y}$
and $\mathbf e_2=\frac{2\pi}{\Lambda_b}\mathbf{\hat x}-\frac{2\pi}{\sqrt3\Lambda_b}\mathbf{\hat y}$.  The parameters 
$\Lambda_s$ and $\Lambda_b$ denote the period of the stripe and bubble phases, respectively.  The results shown in Fig.~\ref{bs} indicate  
interesting differences from electrons in higher LLs \cite{Koulakov} and also from composite fermions in the lowest Landau level \cite{Lee}.
The stripe phase is favored in an intermediate range $0.12\lesssim\nu'\lesssim0.4$; the bubble crystal is better for $\nu'\lesssim 0.12$; and the two are very competitive 
close to half-filling $0.4\lesssim\nu'<0.5$.  
The periods are $8.5<\Lambda_b<10$ and $\Lambda_s\approx 8-9$, apart from small $\nu'$ where the stripe phase is irrelevant.

\begin{figure}[!htbp]
\begin{center}
\includegraphics[width=\columnwidth, keepaspectratio]{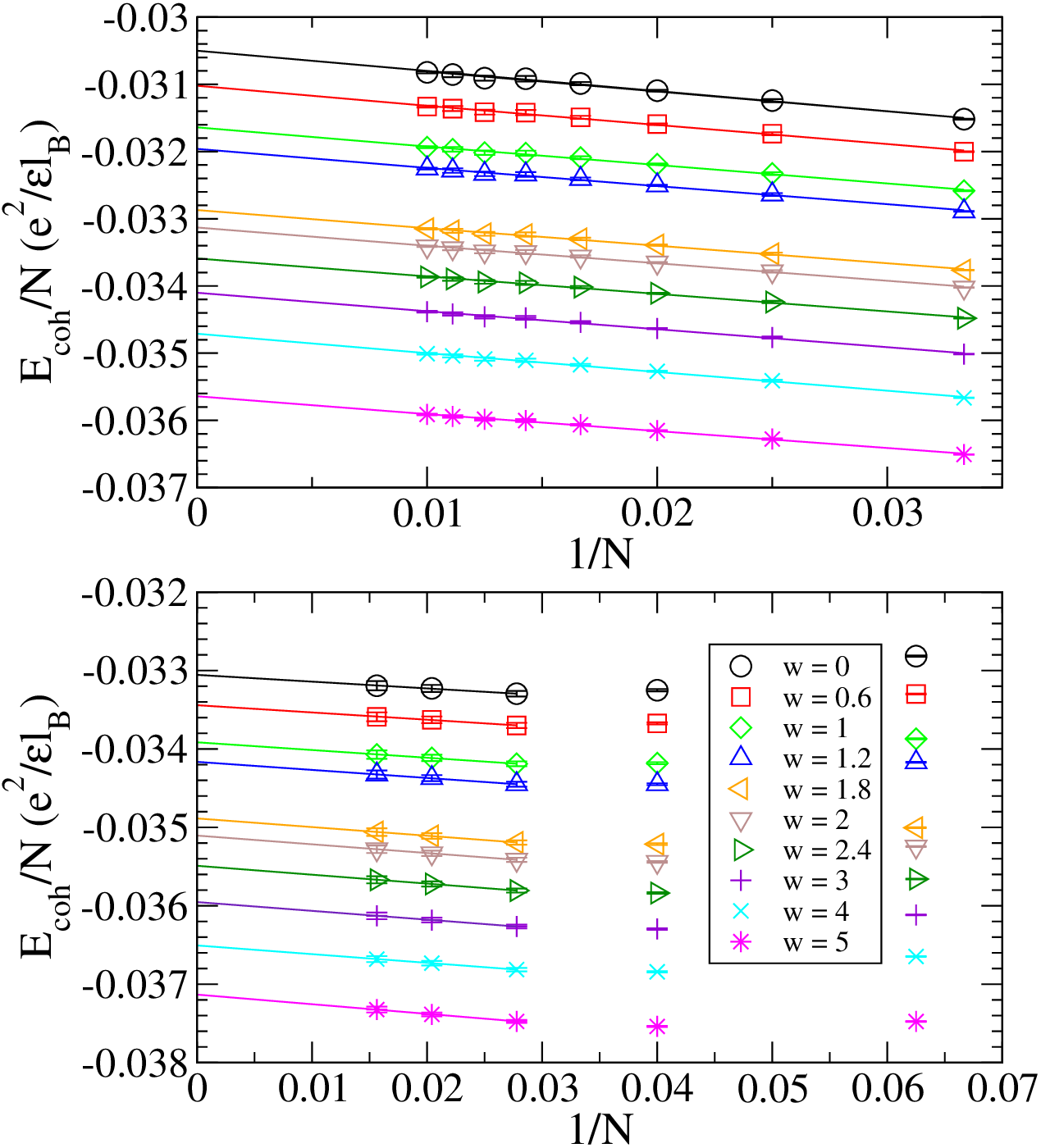}
\end{center}
\caption{\label{vari} (Color online)
Thermodynamic extrapolation of the cohesive energy of the Pfaffian and the CF Fermi sea wave functions.
\CORR{The solid line extends to the right up to the smallest system included the linear fitting ($N=30$ and 36, respectively).}
}
\end{figure}

The sharp short-range repulsion between composite fermions (Fig.~\ref{ppcomp}) suggests the possibility of further vortex attachment to
account for correlations between them. We consider the Pfaffian wave function\cite{theory5p2}, 
\begin{equation}
\label{eqpf}
\Psi^{\text{Pf}}=\prod_{i<j}(u_iv_j-v_iu_j)^2\text{Pf}\left(\frac{1}{u_iv_j-v_iu_j}\right),
\end{equation}
which describes an incompressible $p$-wave paired state of composite fermions, and the compressible CF Fermi sea,
\begin{equation}
\Psi^{\text{CFFS}}={\cal P_{\rm LLL}}\prod_{i<j}(u_iv_j-v_iu_j)^2\Phi_{\rm FS}.
\end{equation}
\CORR{For a comparison with the charge-density-wave states we calculate the cohesive energy of $\Psi^{\text{Pf}}$ and $\Psi^{\text{CFFS}}$ by
subtracting the energy of the uniform Hartree-Fock state (Eq.~(\ref{uniform})).}
The cohesive energies of $\Psi^{\text{Pf}}$ and $\Psi^{\text{CFFS}}$ are shown in Fig.~\ref{vari}.
\CORR{The extrapolation is based on $N=30-100$ particle systems for the Pfaffian, and $N=36,49,64$ for the Fermi sea.}
In the thermodynamical limit $N\to\infty$ both the Pfaffian and the CFFS states are energetically favored over the charge-density-wave states (Table \ref{varienergy}).
The Pfaffian wave function has higher energy for the whole range of transverse thickness studied, making Pfaffian-CF pairing unlikely as a mechanism for incompressibility at $2+3/8$.

\begin{table}[htb]
\begin{center}
\begin{tabular}{l|l|l}
\hline\hline
$w/\ell_B$ & $E_{\text{coh}}$ of $\Psi^{\text{Pf}}$ &  $E_{\text{coh}}$ of $\Psi^{\text{FS}}$ \\
\hline
0    & -0.03050(2) & -0.0331(1)\\
0.6  & -0.03102(2) & -0.0335(1)\\
1    & -0.03164(2) & -0.0339(1)\\
1.2  & -0.03196(2) & -0.0342(1)\\
1.8  & -0.03287(2) & -0.0349(1)\\
2    & -0.03313(2) & -0.0351(1)\\
2.4  & -0.03359(2) & -0.0355(1)\\
3    & -0.03410(2) & -0.0360(1)\\
4    & -0.03471(2) & -0.0365(1)\\
5    & -0.03564(1) & -0.0372(1)\\
\hline\hline
\end{tabular}
\end{center}
\caption{\label{varienergy}
The cohesive energy of the Pfaffian and the CF Fermi sea wave functions.
}
\end{table}

\section{Exact diagonalization}
\label{exact}

We have confirmed the above conclusion by performing exact diagonalization for fermions at $\nu'=1/2$ interacting with the potential given in
Eq.~(\ref{effective}).  Exact diagonalization at $2Q=2N-3$ for $N=8,10$ shows the ground state to be nonuniform,
consistent with the variational study ruling out the Pfaffian wave function.
However, we find that the quantum number of the ground state at $2Q=2N-2$ agrees with the prediction of the CF
theory for $N=4-11$ (see Table \ref{fermil}).
%%%% (two filled $\Lambda$ levels) to $N=11$ (three filled $\Lambda$ levels plus two CF quasiparticles at max.\ separation).
\CORR{The same quantum numbers result for the half-filled lowest Landau level, where the CF Fermi sea state is well established\cite{HLR,FSexp}.}

\begin{table}[htb]
\begin{center}
\begin{tabular}{l|l|l|l}
\hline\hline
$N$ & $Q$ & $L$ & CF interpretation \\
\hline
4  &  6 & 0 & $\Phi_2$ \\
5  &  8 & 2 & $\Phi_2$ $+$ a CF quasiparticle with $L=2$ \\
6  & 10 & 3 & $\Phi_2$ $+$ two CF quasiparticles at maximum separation \\ 
7  & 12 & 3 & $\Phi_3$ $+$ two CF quasiholes \\
8  & 14 & 2 & $\Phi_3$ $+$ a single CF quasihole \\
9  & 16 & 0 & $\Phi_3$  \\
10 & 18 & 3 & $\Phi_3$ $+$ a CF quasiparticle with $L=3$ \\
11 & 20 & 5 & $\Phi_3$ $+$ two CF quasiparticles at maximum separation \\
\hline\hline
\end{tabular}
\end{center}
\caption{\label{fermil}
The ground state angular momentum for $2Q=2N-2$ on the sphere, along with its 
interpretation.  The state with $n$ filled $\Lambda$ levels is denoted by $\Phi_n$.}
\end{table}

\section{Composite fermion diagonalization}
\label{cfdiag}

\begin{figure}[tbp]
\begin{center}
\includegraphics[width=\columnwidth, keepaspectratio]{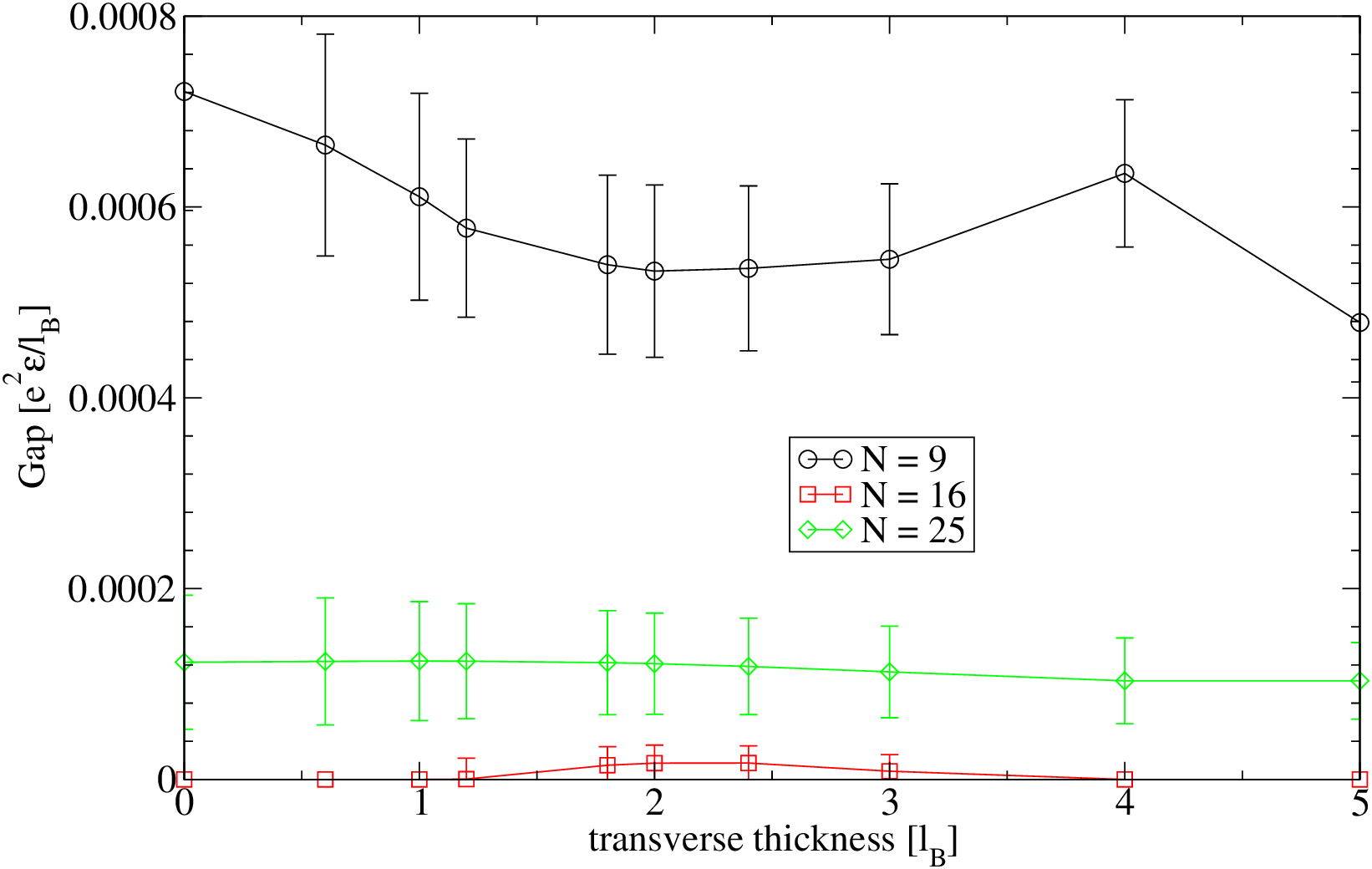}
\end{center}
\caption{\label{gap} (Color online)
The collective mode gap at $\nu=2+3/8$ from second-order CF diagonalization for $N=9,16,25$ particles.
\CORR{The bars indicate the statistical uncertainty arising from Monte Carlo sampling.}
}
\end{figure}

While the noninteracting CF Fermi sea is compressible, the possibility that the residual inter-CF interaction may give rise to
incompressibility cannot be \emph{a priori} excluded.
An investigation of this physics requires larger systems than can be addressed in exact diagonalization.
We have studied this problem by a perturbative process called CF diagonalization\cite{Mandal,paper3}.
(Notice that we form $^4$CFs out of $^2$CFs, and our starting point is a residual inter-$^2$CF interaction $V^{\text{eff}}$.) 
The wave functions for noninteracting composite fermions, for ground as well as excited states, can be constructed by analogy with
the system of noninteracting electrons at an effective filling\cite{JainKamilla}.
These are given by 
\begin{equation}
\Psi_Q = P_{\rm LLL}\prod_{i<j}(u_iv_j-v_iu_j)^2 \Phi_{Q^\ast}
\end{equation}
(here, $\Phi_{Q^\ast}$ is the wave function for the ground or excited state at $Q^*=0$),
and form bands separated by an effective CF cyclotron energy.
At the $n$-th order CF diagonalization, we diagonalize the \CORR{residual inter-CF} interaction in the truncated space of correlated
wave functions of the lowest $n+1$ bands, using the Metropolis Monte Carlo method\cite{Mandal}.
It is necessary to go to at least second order to allow hybridization of the uniform (with orbital angular momentum $L=0$) state with ``excited" bands.
At the second order, the ground state has $L=0$, with a very small gap which depends only weakly on the layer thickness, as shown in Fig.~\ref{gap}.
The enhancement of the gap from $N=16$ to $N=25$ suggests possible establishment of incompressibility due to the residual interaction between composite fermions.
While our data in the $N\le25$ range, unfortunately, do not allow for an extrapolation of the gap to the thermodynamical limit,
it is clear that the gap is extremely small.

\section{Conclusion}
\label{concl}

In conclusion, modeling the system at $\nu=2+3/8$ as filling factor $3/2$ of fully spin polarized composite fermions
in the second electronic Landau level, we have considered many possible structures by several methods.
Our study suggests the possibility of a very delicate FQHE here due to residual interactions between composite fermions,
but with a state distinct from the Pfaffian state. \CORR{At the end, we note that even though our CF diagonalization approach gives the low energy spectrum, it does not provide a simple wave function for the ground state, which has often been very useful in achieving a physical understanding of the physics of a FQHE state.}

\section*{Acknowledgement}

We thank the Center for Scientic Computing at J.~W.~Goethe-Universit\"at for computing time on Cluster III.

\newcommand{\PRL}{Phys.\ Rev.\ Lett.}
\newcommand{\PRB}{Phys.\ Rev.\ B}
\newcommand{\NPB}{Nucl.\ Phys.\ B}


\begin{thebibliography}{99}

\bibitem{Tsui} D.~C.~Tsui, H.~L.~Stormer, and A.~C.~Gossard, \PRL\ {\bf 48}, 1559 (1982).

\bibitem{Jain} J.~K.~Jain, \PRL\ {\bf 63}, 199 (1989).

\bibitem{exp5p2} R.\ Willett, J.P.\ Eisenstein, H.L.\ Stormer, D.C.\ Tsui, A.C.\ Gossard, and J.H.\ English, \PRL\ \textbf{59}, 1776 (1987); W. Pan, J.-S. Xia, V. Shvarts, D. E. Adams, H. L. Stormer, D. C. Tsui, L. N. Pfeiffer, K. W. Baldwin, and K. W.  West, \PRL\ {\bf 83}, 3530 (1999). 

\bibitem{theory5p2} G.\ Moore and N.\ Read, \NPB\ \textbf{360}, 362 (1991);
M.\ Greiter, X.G.\ Wen, and F.\ Wilczek, \PRL\ \textbf{66}, 3205 (1991); \NPB\ \textbf{374}, 567 (1992).

\bibitem{Chang} A. M. Chang, P. Berglund,  D.~C.~Tsui, H.~L.~Stormer, and J. C. M. Hwang, \PRL\ \textbf{53}, 997 (1984).

\bibitem{Pan1} W. Pan, H.L. Stormer, D.C. Tsui, L.N. Pfeiffer, K.W. Baldwin, and K.W. West, Phys. Rev. Lett. {\bf 90}, 016801 (2003). 

\bibitem{Quinn} A. W\'{o}js, K.S. Yi, and J.J. Quinn, Phys. Rev. B {\bf 69}, 205322 (2004).

\bibitem{Pan2} W. Pan, J. S. Xia, H. L. Stormer, D. C. Tsui, C. Vincente, E. D. Adams, N. S. Sullivan, L. N. Pfeiffer, K. W. Baldwin, and K. W. West, Phys. Rev. B {\bf 77}, 075307 (2008);
J. S. Xia, W. Pan, C. Vincente, E. D. Adams, N. S. Sullivan, H. L. Stormer, D. C. Tsui, L. N. Pfeiffer, K. W. Baldwin, and K. W. West, \PRL\ \textbf{93}, 176809 (2004).

\bibitem{Kang} H. C. Choi, W. Kang, S. Das Sarma, L. N. Pfeiffer, and K. W. West, \PRB\ \textbf{77}, 081301(R) (2008).

\bibitem{HLR} V.~Kalmeyer and S.~C.~Zhang, \PRB\ \textbf{46}, R9889 (1992);
B.~I.~Halperin, P.~A.~Lee, and N.~Read, \PRB\ \textbf{47}, 7312 (1993).

\bibitem{FSexp} R.L.~Willett, R. R. Ruel, K. W. West, and L. N. Pfeiffer, \PRL\ {\bf 71}, 3846 (1993);
W.~Kang, H. L. Stormer, L. N. Pfeiffer, K. W. Baldwin, and K. W. West, \PRL\ {\bf 71}, 3850 (1993);
V.J.~Goldman, B. Su, and J. K. Jain, \PRL\ {\bf 72}, 2065 (1994);
J.H.~Smet, D. Weiss, R. H. Blick, G. Lutjering, K. von Klitzing, R. Fleischmann, R. Ketzmerick, and T. Geisel, G. Weimann, \PRL\ {\bf 77}, 2272 (1996).

\bibitem{newodd} K.~Park and J.~K.~Jain, \PRB\ \textbf{62}, R13274 (2000);
C-C.~Chang and J.~K.~Jain, \PRL\ \textbf{92}, 196806 (2004);
A.~Lopez and E.~Fradkin, \PRB\ \textbf{69}, 155322 (2004);
M.~O.~Goerbig, P.~Lederer, and C.~M.~Smith, \PRB\ \textbf{69}, 155324 (2004).

\bibitem{Mandal} S.~S.~Mandal and J.~K.~Jain, \PRB\ \textbf{66}, 155302 (2002).

\bibitem{CFint}
J. K. Jain, R. K. Kamilla, K. Park, and V. W. Scarola, Solid State Commun.\ \textbf{117}, 117 (2002);
S. S. Mandal and J. K. Jain, \textbf{89}, 096801 (2002);
S. S. Mandal, M. R. Peterson, and J. K. Jain, \PRL\ \textbf{90}, 106403 (2003);
C.~T\H oke, M.~R.~Peterson, G.~S.~Jeon, and J.~K.~Jain, \PRB\ \textbf{72}, 125315 (2005).

\bibitem{paper3} C.~T\H oke and J.~K.~Jain, \PRL\ \textbf{96}, 246805 (2006).

\bibitem{Lee} S-Y.~Lee, V.~W.~Scarola, and J.~K.~Jain, \PRL\ \textbf{87}, 256803 (2001);
S-Y.~Lee, V.~W.~Scarola, and J.~K.~Jain, \PRB\ \textbf{66}, 085336 (2002).

\bibitem{Wojs2} A. W\'ojs and J. J. Quinn, Physica E \textbf{40}, 967 (2008).


\bibitem{Shi} C. Shi, S. Jolad, N. Regnault, and J. K. Jain, \PRB\ \textbf{77}, 155127 (2008).

\bibitem{Haldane} F.~D.~M.~Haldane, \PRL\ \textbf{51}, 605 (1983); and
in \textit{The Quantum Hall Effect}, ed.\ by R.~E.~Prange and S.~M.~Girvin (Springer, New York, 1987);
G.~Fano, F.~Ortolani, and E.~Colombo, \PRB\ \textbf{34}, 2670 (1986).

\bibitem{murthy} G. Murthy and R. Shankar, Rev.\ Mod.\ Phys.\ \textbf{75}, 1101 (2003).

\bibitem{Ambru} N.~d'Ambrumenil and A.~M.~Reynolds, J.\ Phys.\ C: Solid State Phys.\ \textbf{21}, 119 (1988).

\bibitem{Wojs} A. W\'ojs and J. J. Quinn, \PRB\ \textbf{61}, 2846 (2000).

\bibitem{Wu} T.~T.~Wu and C.~N.~Yang, \NPB\ {\bf{107}}, 365 (1976).

\bibitem{Koulakov} A. A. Koulakov, M. M. Fogler, B. I. Shklovskii, \PRL\ \textbf{76}, 499 (1996);
M. M. Fogler, A. A. Koulakov, B. I. Shklovskii, \PRB\ \textbf{54}, 1853 (1996).

\bibitem{JainKamilla} J.K.\ Jain and R.K.\ Kamilla, Int. J. Mod. Phys.\ {\bf{B11}}, 2621 (1997); \PRB\ \textbf{55}, R4895 (1997).

\bibitem{liquidcrystal} V. Oganesyan, S. A. Kivelson, and E. Fradkin, \PRB\ \textbf{64}, 195109 (2001);
O. Ciftja and C. Wexler, \PRB\ \textbf{65}, 205307 (2002);
C. Wexler and O. Ciftja, Int.\ J.\ Mod.\ Phys.\ \textbf{20}, 747 (2006);
Q. M. Doan and E. Manousakis, \PRB\ \textbf{75}, 195433 (2007).



%%%%%%%%%%%%%%%%%%%%%%%%

%%%\bibitem{Dean} C. R. Dean, B. A. Piot, P. Hayden, S. Das Sarma, G. Gervais, L. N. Pfeiffer, and K. W. West, \PRL\ \textbf{100}, 146803 (2008).

\end{thebibliography}
\end{document}